\documentclass[useAMS,usenatbib]{mn2e}
\usepackage{epsfig}

\newcommand{\target}{V404~Cyg}

\newcommand{\HST}{\textit{HST}}

\title[The Quiescent SED of \target]{The Quiescent Spectral Energy Distribution of \target}
\author[R. I. Hynes et al.]{R. I. Hynes$^{1}$\thanks{E-mail:
rih@phys.lsu.edu}, C. K. Bradley$^1$, 
M. Rupen$^2$, 
E. Gallo$^3$\thanks{Hubble Fellow},
R. P. Fender$^4$, 
J. Casares$^5$, 
\newauthor 
C. Zurita$^5$\\
$^{1}$Department of Physics and Astronomy, Louisiana State
University, Baton Rouge, Louisiana 70803, USA\\
$^2$National Radio Astronomy Observatory, Array
  Operations Center, 1003 Lopezville Road, Socorro, NM 87801, USA\\
$^3$Physics Department, Broida Hall, University of California, Santa Barbara, CA 93106, USA\\
$^{4}$School of Physics and Astronomy, 
The University of Southampton, Southampton, SO17 1BJ, UK\\
$^{5}$Instituto de Astrof\'\i{}sica de Canarias, 38200 La Laguna,
Tenerife, Spain}

\begin{document}

\date{Accepted ??. Received ??; in original form ??}
      
\pagerange{\pageref{firstpage}--\pageref{lastpage}} \pubyear{2009}

\maketitle

\label{firstpage}

\begin{abstract}

  We present a multiwavelength study of the black hole X-ray binary
  \target\ in quiescence, focusing upon the spectral energy
  distribution (SED).  Radio, optical, UV, and X-ray coverage is
  simultaneous.  We supplement the SED with additional
  non-simultaneous data in the optical through infrared where
  necessary.  The compiled SED is the most complete available for
  this, the X-ray and radio brightest quiescent black hole system.  We
  find no need for a substantial contribution from accretion light
  from the near-UV to the near-IR, and in particular the weak UV
  emission constrains published spectral models for \target.  We
  confirm that no plausible companion spectrum and interstellar
  extinction can fully explain the mid-IR, however, and an IR excess
  from a jet or cool disc appears to be required.  The X-ray spectrum
  is consistent with a $\Gamma \sim 2$ power-law as found by all other
  studies to date.  There is no evidence for any variation in the
  hardness over a range of a factor of 10 in luminosity.  The radio
  flux is consistent with a flat spectrum (in $f_{\nu}$).  The break
  frequency between a flat and optically thin spectrum most likely
  occurs in the mid or far-IR, but is not strongly constrained by
  these data.  We find the radio to be substantially variable but with
  no clear correlation with X-ray variability.

\end{abstract}


\section{Introduction}

The question of how accretion onto a black hole occurs has been of
great and ongoing interest. Observations of accreting sources provide
a test of astrophysics in the strong gravity regime and will
potentially yield a signature of the presence of an event horizon
around black holes.  Recent advances in X-ray technology have revealed
that accretion can occur at extremely low rates across the range of
black hole masses from stellar remnants to supermassive black holes at
the centre of our own and other galaxies.  However, the nature of that
low-level accretion remains unclear (see e.g.,
\citealt{Narayan:2008a}).  Accretion flows at low luminosities do
appear to be radiatively inefficient, but we have yet to determine to
what extent that arises from advection of hot gas through the event
horizon, and what role jets may play in carrying away energy
\citep{Fender:2003a}.  These low-level accreting black holes emit
across the electromagnetic spectrum from radio to X-rays, so
multiwavelength studies can be used to disentangle different sources
from different regions of the inflow, outflow, or jet, and to
establish causal connections between behavior in these regions.

Among the stellar mass black hole population, the best studied
quiescent system for this purpose has been \target, which contains a
compact object with a mass function of $6.08\pm0.06$\,M$_{\odot}$
\citep{Casares:1994a} -- one of the most secure stellar mass black
hole mass determinations we have.  \target\ has the longest orbital
period of the black hole X-ray transients (BHXRTs), at 6.5\,days, with
a K0\,IV companion star \citep{Casares:1993a}.  It is also the most
luminous stellar mass black hole in quiescence with $L_{\rm X} \sim
10^{33}-10^{34}$\,erg\,s$^{-1}$, $\sim1000\times$ brighter than the
prototypical BHXRT, A\,0620--00 \citep{Garcia:2001a}.  These extremes
are unlikely to be coincidental and the high $L_{\rm X}$ may be a
consequence of the long orbital period, the evolved donor star, and/or
the high mass of the black hole.

Attempts to relate observations of quiescent black hole binaries to
theoretical models of accretion flows have focused upon fitting the
broad-band spectral energy distribution, ideally with coverage from
radio to X-ray energies.  \target\ has been a prime target due to its
X-ray brightness \citep{Narayan:1996a,Narayan:1997a,Quataert:1999a}.
These studies mainly focused on pure accretion models and only
considered the optical to X-ray regime, as radio data were limited or
non-existent.  Attempts to apply advective models to the more luminous
hard state (e.g.\ \citealt{Esin:2001a}) brought home the fact that
these models could not explain the ubiquitous radio emission in this
state, that was sometimes seen to extend as far as the near-IR band.
A more complete description of the low-state SED appears to require a
coupled accretion plus outflow model (\citealt{Markoff:2001a};
\citealt{Yuan:2005a}).  More recent studies have revealed significant
radio emission in quiescence too, from \target\ \citep{Gallo:2005a},
and from A\,0620--00 \citep{Gallo:2006a} providing support to the idea
that an outflow also plays a significant role in quiescence.

To facilitate comparison between developing models and data we present
here the most complete SED available for \target.  It is primarily
based on a simultaneous campaign performed in 2003 using {\it
  Chandra}, {\it HST}, and several optical and radio observatories.
We supplement the coverage with non-simultaneous optical and near-IR
data from \citet{Casares:1993a} and near to mid-IR {\it Spitzer} data
from \citet{Muno:2006a} and \citet{Gallo:2007a}. Previous results from
this campaign were presented by \citet{Hynes:2004a}; see also
\citet{Corbel:2008a}.

\section{Observations}

\subsection{{\it HST} ultraviolet data}

Ultraviolet photometry used the Advanced Camera for Surveys (ACS;
\citealt{Boffi:2007a}) onboard the {\em Hubble Space Telescope (HST)}.
The observations performed are summarised in Table~\ref{HSTLogTable}.
V404~Cyg is a faint UV source and reddened, so five orbits were used
covering the F250W and F330W filters.

\begin{table}
\caption{Log of {\it HST} observations}
\label{HSTLogTable}
\begin{tabular}{llcl}
\hline
\noalign{\smallskip}
Date & UT Start& Duration (s) & Filter \\
\noalign{\smallskip}
\hline
\noalign{\smallskip}
2003 July 28 & 19:13:10 & 200 & F330W \\ 
2003 July 28 & 19:19:08 & 400 & F250W \\ 
2003 July 28 & 19:28:22 & 200 & F330W \\ 
2003 July 28 & 19:34:20 & 400 & F250W \\ 
2003 July 28 & 20:28:46 & 200 & F330W \\ 
2003 July 28 & 20:34:44 & 400 & F250W \\ 
2003 July 28 & 21:09:32 & 200 & F330W \\ 
2003 July 28 & 22:00:19 & 400 & F250W \\ 
2003 July 28 & 22:09:33 & 200 & F330W \\ 
2003 July 28 & 22:25:28 & 400 & F250W \\ 
2003 July 28 & 22:34:42 & 200 & F330W \\ 
2003 July 28 & 22:40:40 & 400 & F250W \\ 
2003 July 28 & 23:38:31 & 980 & F250W \\ 
2003 July 29 & 01:18:45 & 980 & F250W \\ 
2003 July 29 & 04:26:38 & 980 & F250W \\
\noalign{\smallskip}
\hline
\end{tabular}
\end{table}

We used standard ACS techniques to reduce the data.  Individual images
had been pre-processed to the flat-fielded stage using the automatic
pipeline, CALACS; we found no need to refine this reduction.  We then
combined all of the images offline into a geometrically corrected
master image using Multidrizzle \citep{Koekemoer:2002a} with standard
settings.  We found excellent registration between individual images,
so no need to improve the shifts applied.

We performed photometry using the IDL/AstroLib aperture photometry
routine {\sc aper}.  As V404~Cyg is faint we used a relatively small
aperture to perform photometry, of radius 0.125\arcsec, with
background defined by a 2.5\arcsec\ annulus.  No stars in the field
were bright enough to determine per-field aperture corrections, so we
use the tabulated values from \citet{Sirianni:2005a}.  This aperture
collects about 75\,\%\ of the total light, and almost all of the sharp
core of the point spread function.  We measured the position of the
target in the F330W bandpass and then fixed it for F250W, after
verifying that it was consistently placed in the two.  We note in
passing that we do detect both \target\ and the nearby contaminating
star.  We find an offset of the contaminating star relative to
\target\ of $\Delta\alpha=+0.005$\,sec ($=0.06$\arcsec) and
$\Delta\delta=1.43$\arcsec, an improvement on the relative astrometry
of the stars compared to that reported by \citet{Casares:1993a}.

A significant source of systematic uncertainty in ACS photometry is
charge transfer inefficiency (CTI) due to radiation damage to the CCDs
(\citealt{Riess:2004a}; \citealt{Pavlovsky:2005a}), although
fortunately the observations were performed early enough in the
lifetime of ACS that CTI is unlikely to be a severe problem.  The
effect is worst at low light levels, so maximised for faint UV
sources.  CTI is also least well calibrated for these cases, and the
correction prescriptions provided by \citet{Pavlovsky:2005a} formally
diverge for negligible sky background or source counts.  We face both
problems with these data.  Sky backgrounds in the near-UV are
extremely low, dominated by zodiacal light and earthshine in the F330W
bandpass, and geo-coronal emission at shorter wavelengths.  Both were
essentially unmeasurable in our images, and the background is expected
to be dominated by CCD dark current, producing 0.5--2.0\,e$^-$/pixel
for our exposures.  We attempted to estimate the CTI losses in two
ways.  Firstly, we followed the prescription in
\citet{Pavlovsky:2005a}, except that we used the estimated background
(including the dominant dark current term) from the ACS Exposure Time
Calculator.  This prescription predicted losses of $\sim 5$\,\%.
Based on the actual data shown by \citet{Riess:2004a} however, this
prescription appears to overestimate the CTI losses for the faintest
sources.  For the low background case (0.5--1.5\,e$^-$/pixel) at
source brightnesses up to a few hundred electrons, \citet{Riess:2004a}
found losses of around 0.035\,mag at maximum distance from the readout
amplifier.  If we rescale this to the times of our observations and
positions of the source we expect losses of about 2.5\,\%.  We use
these latter estimates to correct our measured source brightnesses and
assign an additional error estimate of 2.5\,\%\ to account for the
uncertainty in applying CTI corrections.  The uncertainties in the CTI
corrections are much smaller than the statistical uncertainties in the
measured fluxes, so this is not the dominant source of uncertainty.

We tabulate our photometric results in Table~\ref{HSTPhotTable}.
Since we are measuring counts across a broad bandpass rather than
monochromatic fluxes, we tabulate the data in two ways; the corrected
number of electrons per second, and the average flux per unit
wavelength using photometric zero-points estimated by the ACS pipeline.
All have been corrected for CTI, and to a nominal infinite aperture.

\begin{table}
\caption{Results of {\it HST} photometry}
\label{HSTPhotTable}\begin{tabular}{lcc}
\hline
\noalign{\smallskip}
Band  & Count rate        & Flux \\
      & (s$^{-1}$)          & (erg\,cm$^{-2}$\,s$^{-1}$\,\AA$^{-1}$) \\
\noalign{\smallskip}
\hline
\noalign{\smallskip}
\multicolumn{3}{l}{\em Observed} \\
F250W & $0.109 \pm 0.042$ & $(0.52 \pm 0.20) \times10^{-18}$ \\
F330W & $1.06  \pm 0.14$  & $(2.37 \pm 0.30) \times10^{-18}$ \\
\noalign{\smallskip}
\multicolumn{3}{l}{\em Synthetic K0\,{\sc iii}} \\
F250W & 0.073 & \\
F330W & 0.57 & \\
\noalign{\smallskip}
\multicolumn{3}{l}{\em Synthetic K0\,{\sc iii}, truncated at 4200\,\AA} \\
F250W & 0.024 & \\
F330W & 0.57  & \\
\noalign{\smallskip}
\hline
\end{tabular}
\end{table}

\subsection{IAC80 photometry}

We obtained simultaneous $R$ band observations using the Thomson CCD
camera at the 0.8\,m IAC80 from 20:50--05:51 UT on the night of 2003
July 28/29.  We used 300\,s exposures with 52\,s readout time between
them and bias-corrected and flat-fielded all images in the standard
way using IRAF\footnote{IRAF is distributed by the National Optical
  Astronomy Observatories, which are operated by the Association of
  Universities for Research in Astronomy, Inc., under cooperative
  agreement with the National Science Foundation.}.  Seeing conditions
were not good enough to cleanly separate the contribution of the
target and its nearby (1.4\arcsec) line-of-sight star
\citep{Udalski:1991a} so we applied straightforward aperture
photometry using a large aperture of 3.5\arcsec which adds the flux
from both stars as described by \citet{Zurita:2004a}.  We determine an
average magnitude $R=16.63\pm0.01$ in acceptably close agreement with
\citet{Casares:1993a}, indicating little difference in the optical
brightness of the system compared to earlier data.

\subsection{WHT optical spectroscopy}

To reconstruct the optical SED of the source we use the WHT data from
the program.  This covered a larger range of wavelengths than the
Gemini data, and was simultaneous with the \HST\ observations.  These
observations used the ISIS dual-arm spectrograph on the WHT.  To
maximise efficiency and minimise readout time and noise, we used the
single red-arm mode with the R316R grating and MARCONI2 CCD.  We set
exposure times to 200\,s, with $\sim17$\,s dead-time between
exposures.  To maximise photometric accuracy, we used a 4\arcsec\
slit, so our spectral resolution was determined by the seeing (median
$\sim1.3$\arcsec), and was typically $\sim5.5$\,\AA\
(250\,km\,s$^{-1}$).  We performed bias correction and flat fielding
using standard IRAF techniques.  The slit was oriented in the same was
as in our previous observations and covered the same comparison star.
We extracted spectra of both of these stars, and the nearby blended
star, with the same techniques previously described
(\citealt{Hynes:2002b}; \citealt{Hynes:2002a}).  We performed
wavelength calibration relative to a single observation of a CuNe/CuAr
lamp and corrected for time-dependent variations in the wavelength
calibration using Telluric absorption features.  We calibrated the
on-slit comparison star relative to Kopff 27 \citep{Stone:1977a}, and
then calibrated all spectra of \target\ relative to this on-slit
comparison.

To compare the WHT data with that from the IAC80 we perform synthetic
photometry convolving the spectrum with a $R$ bandpass.  We found an
offset between the two, most likely due to systematic errors in the
WHT calibration.  Since the IAC80 photometry agrees well with earlier
measures we assume that this calibration is correct and rescaled the
WHT data up by a factor of 1.3 to agree with the IAC80 $R$ band data.

\subsection{Gemini data}

We used additional optical spectroscopy obtained with the R831 grating
and standard EEV CCDs of the GMOS spectrograph on Gemini-N to extend
our continuum lightcurve. Exposure times were 40\,s and we binned and
windowed the images to reduce the dead-time between exposures to
12\,s.  With a 5\arcsec\ slit and 1.1\arcsec\ median seeing we
obtained a spectral resolution of 5.0\,\AA\ (230\,km\,s$^{-1}$).  We
performed data reduction, spectral extraction, and wavelength and flux
calibration in the same way as for the WHT data.  Wavelength
calibration used a CuAr lamp, and we performed flux calibration
relative to the same on-slit comparison star as used for the WHT
observations, and applied the same rescaling to ensure consistency
with IAC80 photometry.

\subsection{{\it Chandra} X-ray data}

We use data taken from two recent Chandra observations of V404 Cyg,
one on 2000 April 26 for 10,295\,s (\citealt{Garcia:2001a},
\citealt{Kong:2002a}), and another on 2003 July 28/29 for a total of
61,200\,s \citep{Hynes:2004a}. The first observation used 1/4
sub-array mode, with time resolution of 1.14\,s; while the second used
1/8 sub-array mode, with a resolution of 0.4\,s. In the original set,
1587 counts were reported, while the second had 1941 counts. We
reanalyzed the standard pipeline-processed level-2 data from both data
sets in CIAO v3.2.  The extraction aperture was set to a radius of
6-pixels and we retained only events between 0.3--7.0\,keV to reduce
the background.  The background can be particularly significant for
the S3 chip, so we used large background regions with a 48-pixel
radius.  The background produced approximately 4.6 counts in the
source aperture for the 2003 observation, and 0.2 counts in the 2000
observation.

\subsection{VLA radio data}

VLA data were obtained in the A configuration from 29 July 00:20 UT
until 14:15 UT.  The data were reduced and each time bin imaged using
standard procedures within AIPS.  Absolute flux calibration was 
obtained relative to 3C~286.

\subsection{WSRT radio data}

\begin{figure}
\includegraphics[scale=0.45]{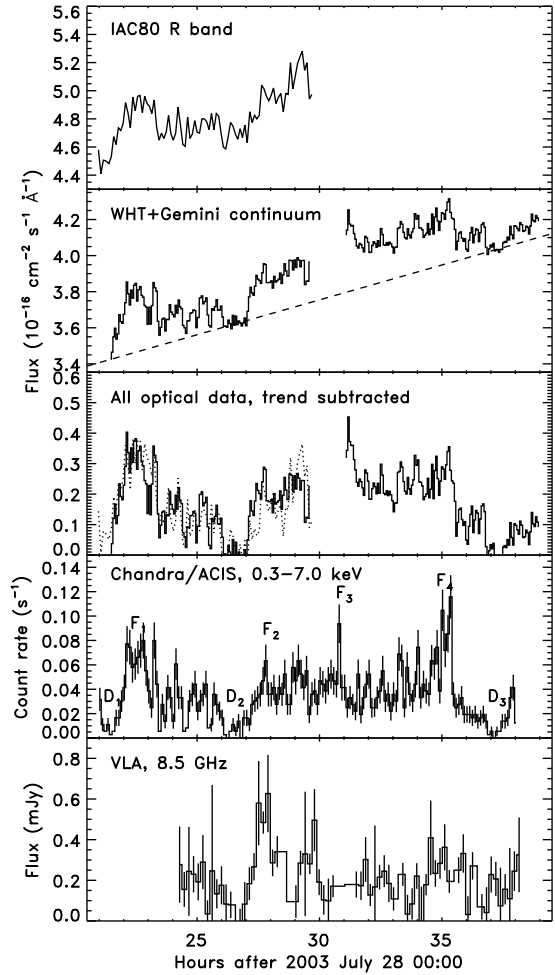}
\caption{Optical, X-ray, and radio lightcurves of \target\ during our
  simultaneous observations.  The upper panels show the full flux
  observed by IAC80 photometry and WHT and Gemini continuum
  spectroscopy.  The dashed line in the second panel is a linear fit
  to the lower-envelope.  The third panel shows the WHT and Gemini
  continuum data after subtraction of this lower-envelope.  The
  photometric data has been treated in the same way after
  re-normalising to agree with the WHT data and is shown dotted.  The
  fourth panel shows the X-ray data and the fifth the radio lightcurve.}
\label{OptVarFig}
\end{figure}

The Westerbork Synthesis Radio Telescope (WSRT) observed \target\ from
28 July 17:13 UT till 29 July 05:33 UT, corresponding to the period
D1--F1--D2--F2 in \citet{Hynes:2004a} and labelled in
Fig.~\ref{OptVarFig}.  We used two frequencies: 4.8 and 1.4\,GHz, with
$\sim5$ hours on source per frequency.  The first hour was unusable
because the source was too low on the horizon.  The full 4.8\,GHz
observation (with useful coverage 18:40--05:13 UT) gives a flux
density of $S_{\rm tot} = 0.19 \pm 0.03$\,mJy.  At 1.4\,GHz the
average flux density was $S_{\rm tot, 1.4} = 0.18 \pm 0.08$\,mJy.

\subsection{Published Non-simultaneous Data}

We supplement our dataset with non-simultaneous points from the
literature to fill in gaps in coverage.  The most comprehensive
coverage of the optical-IR SED is from \citet{Casares:1993a}, who
present $BVR$ and $JHK$ photometry from a coordinated campaign
obtained on 1990 June 10.  These points do raise some concern that
they were taken soon after the outburst and may not be fully
quiescent.  There is also a large gap, 13 years, relative to the
recent data.  Nonetheless, we find excellent agreement in $R$ between
our simultaneous IAC80 photometry and the earlier data.  Recent {\em
  Spitzer} data in the mid-IR have also been presented by
\citet{Muno:2006a} and reanalysed by \citet{Gallo:2007a}.  We 
include these 4.5, 8.0, and 24\,$\mu$m photometry in the SED using
Gallo's data which differ significantly only in the 24\,$\mu$m
measurement.


\section{The UV, Optical, and IR Spectral Energy Distribution}

\subsection{Contamination of the $R$ band by H$\alpha$}

Using the WHT spectrum we can estimate the contribution of line
emission to the photometry and correct for this effect.  The strongest
line is H$\alpha$ with an equivalent width of 16\,\AA.  This
translates to a change in $R$ magnitude of 0.01\,mag, and we correct
for this before using these data in the SED.

\citet{Casares:1993a} found a stronger line during 1990--91 with
average equivalent width 38.7\,\AA.  This suggests a larger
contribution to the earlier photometry of $\sim 2$\,\%, and we use
this value to correct the photometry of \citet{Casares:1993a}.  We
note that in both cases this contamination is effectively negligible
in comparison to other uncertainties.

\subsection{Constraints on the optical disc contribution from variability}
\label{FlickeringSection} 

We expect the optical emission in quiescent BHXRTs to be a combination
of light from the disc and light from the companion star.  In the case
of \target, this is overwhelmingly dominated by the companion.
Traditionally measures of the disc fraction use veiling of the
photospheric lines of the companion star and are dependent on having
high resolution spectra at the time of the observation.  They can be
rather sensitive to assumptions about the spectrum of the companion
star (Hynes et al. 2006).  \citet{Casares:1993a} investigated several
methods of fitting the veiling at different wavelengths within the
optical spectrum.  Despite large uncertainties in their results, they
conclude on approximate values of $r_{\rm B}=0.26$, $r_{\rm V}=0.15$,
and $r_{\rm R}=0.09$, where the $r$ values refer to the fractional
contribution of the disc to the total\footnote{\citet{Casares:1993a}
  quote slightly different values as they define the veiling as the
  disc light as a fraction of the stellar light}.  In the near-IR,
\citet{Shahbaz:1996a} put an upper limit on the $K$ band veiling of
$r_{\rm K}<0.14$.

We cannot make a reliable determination of the disc fraction from our
low resolution spectroscopy, but can instead use a different method to
place a lower limit on the disc fraction.  The optical continuum is
significantly variable, and the variability presumably must be
attributed to the disc.  If we assume that the optical continuum
lightcurve consists of three components: companion star, slowly
varying disc light, and flickering disc light, then the fraction of
light due to the flickering component provides a robust lower limit on
the disc fraction.

We show IAC80 photometry together with WHT and Gemini continuum
lightcurves in Fig.~\ref{OptVarFig}.  Our observations were centred
around photometric phase 0.65 using the ephemeris of
\citet{Casares:1994a} implying that the ellipsoidal modulation should
be rising from the deeper minimum at phase 0.5 to the maximum at phase
0.75.  This rising trend and the approximate amplitude of ellipsoidal
modulations expected are consistent with the general upward trend we
observe.  For a long period system such as \target, we can approximate
the ellipsoidal modulation within a night by a linear trend so we
remove the slowly varying light by subtracting a linear fit to the
lower envelope of the continuum lightcurve.  We find that the
remaining flickering component is about 4\,percent of the total light
in both the WHT and Gemini segments of the lightcurve and correlates
very well with the X-ray variations (as also found for the H$\alpha$
emission by \citealt{Hynes:2004a}).  This 4\,\%\ flickering light is
actually not far off of spectroscopic estimates based on veiling near
H$\alpha$ as discussed above.  This makes it plausible that flickering
light source does contribute most of the disc light in the optical.

\subsection{The stellar contribution}

We do not have sufficient information to independently constrain the
stellar contribution to the SED at each wavelength, so opt to instead
model the stellar spectrum based on estimates of the system
parameters.  From \citet{Casares:1994a} we have a spectral type
estimate of G8--K2\,III--V.  \citet{Shahbaz:1994a} estimates companion
star parameters of $M_2=0.7^{+0.3}_{-0.2}$\,M$_{\odot}$ and
$R_2=6.0^{+0.7}_{-0.5}$\,R$_{\odot}$ implying $\log g = 2.76\pm0.08$
and a giant classification.  The SED of a giant star should be the
best approximation for V404~Cyg, although strictly it is a stripped
giant \citep{King:1993a}.  For a K0\,{\sc iii} spectral type we find
$T_{\rm eff}\simeq4570$\,K using the temperature scale derived by
\citet{vanBelle:1999a} for G, K, and M giants.  We model the spectrum
of the companion by interpolating on the grid of model atmosphere
spectra of late-type giants of \citet{Hauschildt:1999a} after
logarithmically re-binning the models to $\Delta \log \nu = 0.01$.

\begin{figure}
\includegraphics[angle=90,scale=0.35]{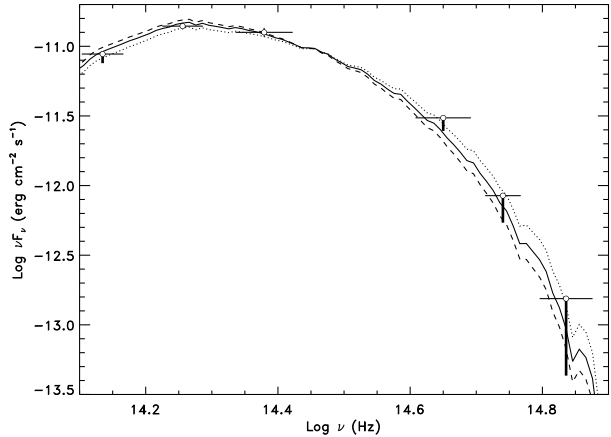}
\caption{Optical to near-IR SED using photometry from
  \citet{Casares:1993a}.  No dereddening has been applied to the data.
  The solid line is the best fit of a K0\,{\sc iii} spectrum reddened
  with $A_{\rm V}=4.04$.  The dotted line corresponds to $A_{\rm
    V}=3.6$ and the dashed line to $A_{\rm V}=4.4$.  The solid lines
  below the $BVR$ and $K$ points are the ranges of plausible companion
  contributions to the SED based on estimates of \citet{Casares:1993a}
  and \citet{Shahbaz:1996a}.}
\label{OptFitFig}
\end{figure}

\citet{Casares:1993a} estimated the interstellar extinction to be
$A_{\rm V}=4.0$ based primarily on comparing the $(B-V)$ colour with
that expected for a K0\,{\sc iii} star after correcting for their
estimates of disc contamination.  We tested this by fitting the model
K0\,{\sc iii} spectrum described above to the full set of BVRJHK
magnitudes (folding the model spectrum with bandpasses to generate
synthetic photometry) constraining the fit to preferably lie within
ranges implied by the disc contamination constraints of
\citet{Casares:1993a} and \citet{Shahbaz:1996a} and shown in
Fig.~\ref{OptFitFig} by vertical bars.  Our fit confirms the earlier
estimate with a formal best fit for $A_{\rm V}=4.04$, shown in
Fig.~\ref{OptFitFig}.  The residuals scatter both above and below the
line indicating that intrinsic variability and/or calibration
uncertainties are playing a role, so we do not attempt to estimate an
error on the extinction but we do show companion spectra reddened by
$A_{\rm V}=3.6$ and $A_{\rm V}=4.4$ to indicate the sensitivity to
extinction.  Extinctions much outside of this range would require
attributing a larger fraction of the light to the disc than has
previously been estimated.

\begin{figure}
\includegraphics[angle=90,scale=0.35]{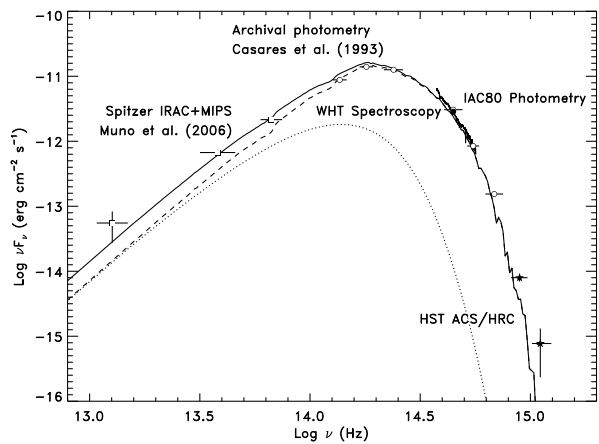}
\caption{UV-Optical-IR SED.  No dereddening has been applied to the
  data.  The short solid line segment and filled symbols indicate new
  data, open symbols mark archival photometry.  The highest frequency
  {\it HST} point has been approximately corrected for red-leak
  (Section~\ref{UVExcessSection}).  The dashed line is a model K0\,III
  SED reddened with $A_{\rm V}=4.04$.  The dotted line is a 2000\,K
  blackbody component adjusted so that the sum of the two (shown with
  a long solid line) fits the short wavelength {\it Spitzer} data.}
\label{OptSEDFig}
\end{figure}

The alternate G8 and K2 possibilities are almost indistinguishable
from K0, so we do not show them.  The K0\,{\sc iii} stellar spectrum
clearly provides a good description of the UV to near-IR data,
consistent with previous estimates that the companion dominates the
light in both the optical and infrared.  We expand the view in
Fig.~\ref{OptSEDFig} to include both near to mid-IR data from {\em
  Spitzer} and near-UV data from {\it HST}.  As found by
\citet{Muno:2006a}, an IR excess does appear to be seen by {\it
  Spitzer}.  A weak UV component also seems to be present, although
this is not as pronounced as seen in other quiescent BHXRTs
(\citealt{McClintock:2000a}; \citealt{McClintock:2003a}; Hynes et
al. in prep.)  The F250W data (the highest frequency) shown in
Fig.~\ref{OptSEDFig} has already been corrected for an estimate of the
red-leak as will be discussed in Section~\ref{UVExcessSection}.

\subsection{The mid-IR excess}

The {\it Spitzer} data clearly lie above a plausible extrapolation of
the companion star spectrum alone (Fig.~\ref{OptSEDFig}).  Given the
insensitivity of the IR spectrum to the temperature of the companion
and interstellar extinction, we cannot adjust the companion spectrum
in any way to fit these IR data.  While these were not simultaneous
with the optical data, we do not expect the continuum brightness of
the companion to vary from epoch to epoch.  Small variations due to
ellipsoidal modulations will be present as a function of orbital
phase, but these are less than $\pm10\,\%$ \citep{Shahbaz:1994a}.

\begin{figure*}
 \begin{center}
  \includegraphics[angle=270,scale=0.3]{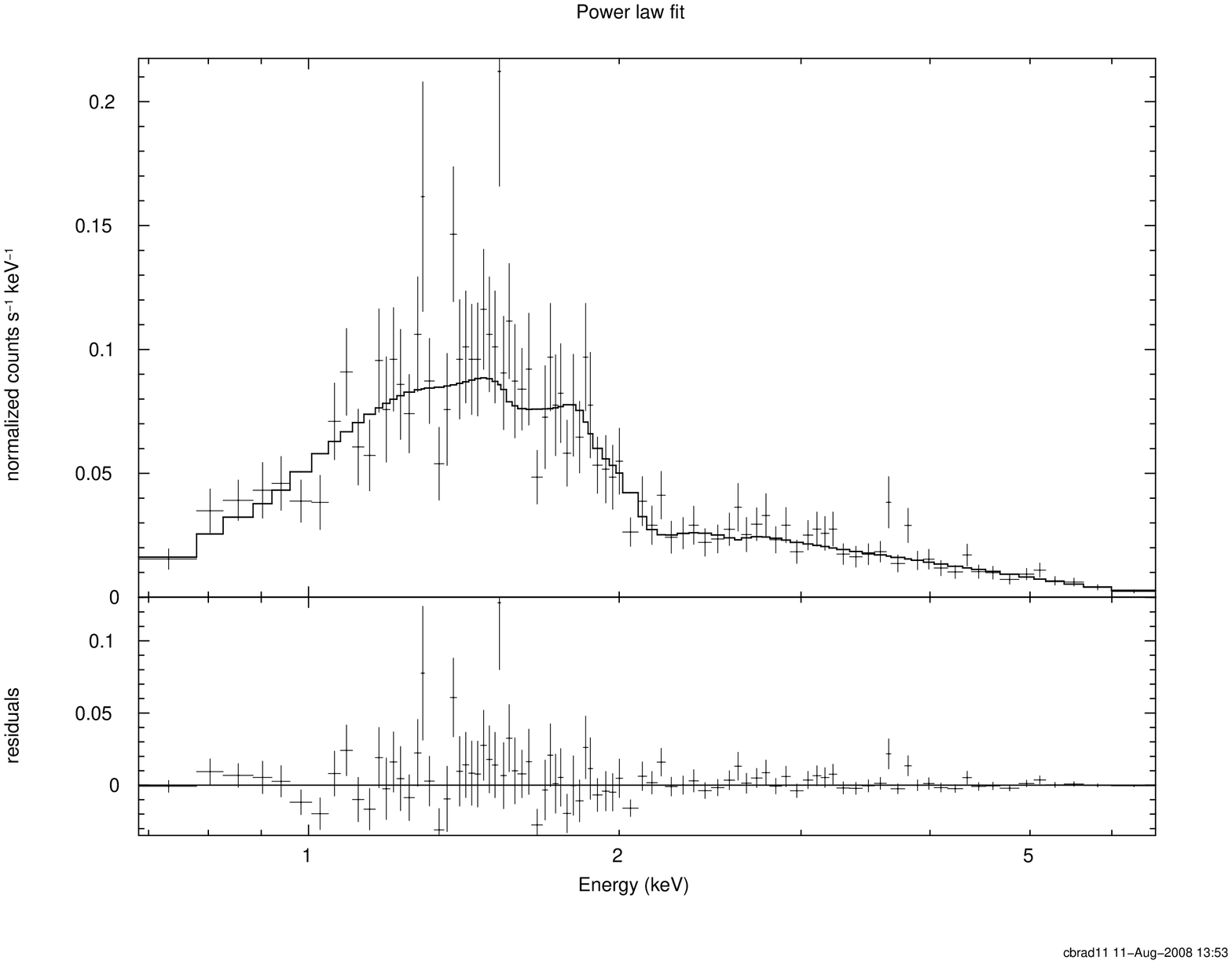}
  \includegraphics[angle=270,scale=0.3]{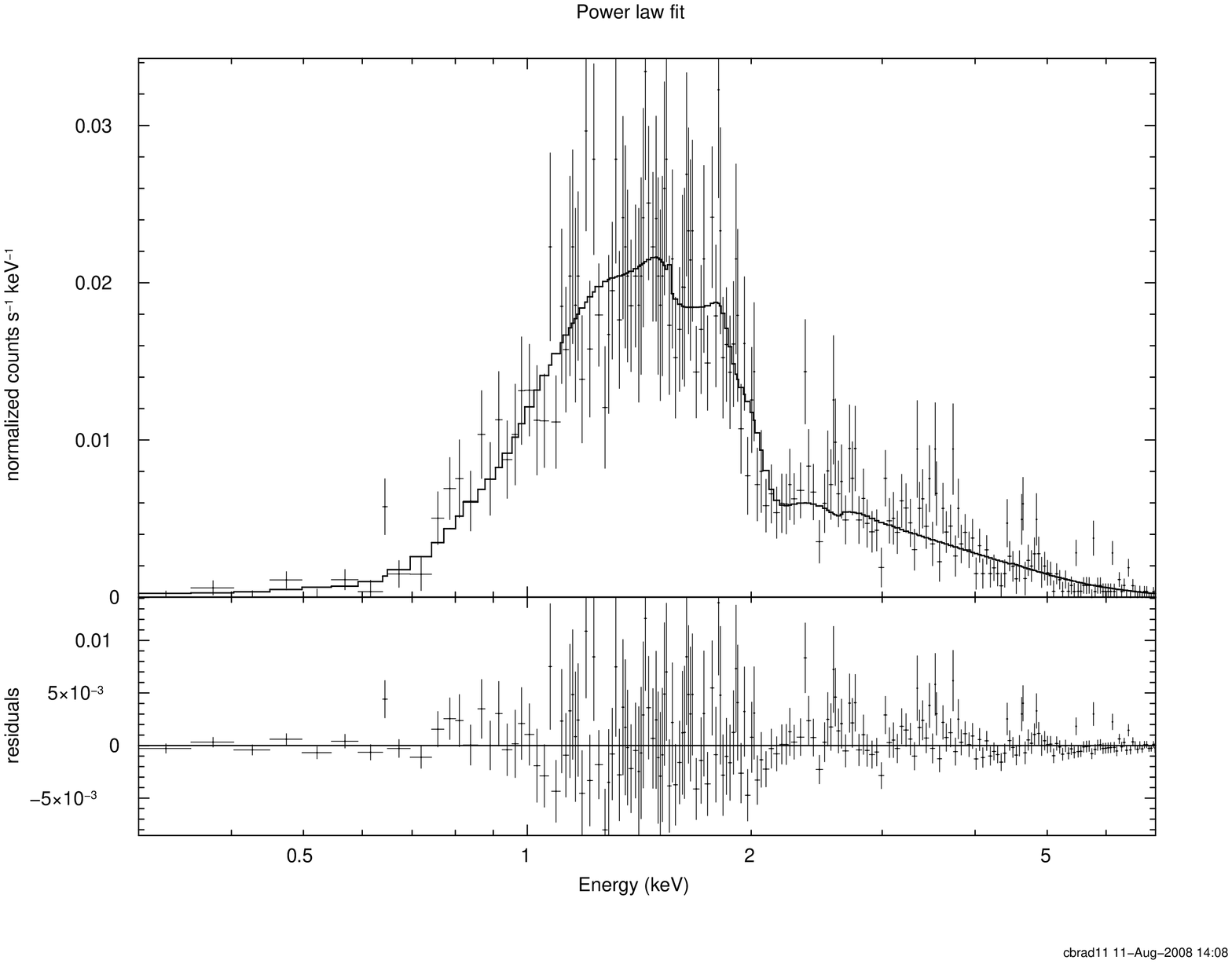}
   \caption{Power-law fits to {\it Chandra} data from 2000 (left) and 2003 (right).}
\label{XSpecFig}
 \end{center}
\end{figure*}

\citet{Muno:2006a} interpreted the mid-IR excess seen in V404~Cyg as
the tail of a Shakura-Sunyaev accretion disc.  This model, however,
predicts a UV excess which is inconsistent with our {\it HST}
observations, exceeding our F330W measurement by an order of magnitude
and the F250W by two orders.  During quiescence instead most of
the disc is expected to be in a cool state, 2000--3000\,K, with a more
isothermal temperature distribution (see \citealt{Hynes:2005a} and
references therein).  The contamination from this material will then
be most pronounced in the mid-IR as it is cooler than the companion
star, but no UV excess will be present.

We use a solid line in Fig.~\ref{OptSEDFig} to show a model including
a 2000\,K blackbody in addition to the companion star spectrum
discussed above.  This can adequately fit both the near-IR to UV SED
and the {\it Spitzer} near-IR data (it is less successful at
24\,$\mu$m).  The blackbody flux is consistent with a disc of radius
$\sim9\times10^{11}$\,cm tilted at an inclination of 56$^{\circ}$
\citep{Shahbaz:1994a} and at a distance of 3.5\,kpc.  For the system
parameters discussed above, this corresponds to about 60\,percent of
the radius of the Roche lobe of the black hole.  This is a very
plausible size for a quiescent accretion disc.

We note that while \citet{Muno:2006a} preferred a Shakura-Sunyaev disc
model for the IR excess in \target\ specifically, in other systems they
invoked a much cooler circum-binary disc.  Such a disc would remain a
possibility in the light of our UV observations as it would not
contribute to the UV flux.

The difficulty in fitting the 24\,$\mu$m point supports a more
controversial interpretation as proposed by \citet{Gallo:2007a}.  The
flux in this band is rather close to an extrapolation of the flat
radio spectrum raising the possibility that the IR excess could be due
to synchrotron jet emission.  The major difficulty in quantifying this
is that the radio is known to be strongly variable
(Section~\ref{RadioSection}) and was not observed simultaneously with
the 24\,$\mu$m point.  We will return to this topic in
Section~\ref{JetSection}.

\subsection{The UV excess}
\label{UVExcessSection}

Our data also provide tenuous evidence for a UV excess.  One possible
explanation for this is red-leaks in the UV bandpasses
\citep{Chiaberge:2007a}.  Although this may make a large contribution
in F250W, the effect is not expected for the F330W filter.  The
estimates for red-leaks from \citet{Chiaberge:2007a} are of no use as
they only consider unreddened stars, so we perform our own
calculations.  In addition to red-leaks, because the spectrum is so
steep in the UV it is hard to gauge the expected flux in a broad-band
without performing synthetic photometry.

To better assess the significance of any UV excess, we therefore use
{\sc synphot} \citep{Laidler:2005a} to estimate count-rates in each
bandpass for our model reddened K0\,{\sc iii} spectrum.  We consider
two spectra, one with the full spectral coverage, and the other
truncated above 4200\,\AA\ to disentangle in-band stellar flux from
the red-leak.  We find that the red-leak is negligible in the F330W
bandpass, as expected, and that the in-band stellar flux can account
for approximately half of the count-rate seen in this band; this is a
little more than a 3$\,\sigma$ excess.  The F250W band appears to have
a large contribution from red-leak, about half of the flux in the
band.  After subtraction of the companion star, both in-band and
red-leak, the residual excess is $0.36\pm0.40$\,s$^{-1}$, so is not a
significant detection of disc light in this band.

We therefore appear to have a significant detection of a UV excess in
the F330W band only, but even this is at the mercy of uncertainties in
the companion spectral classification (which do become a concern in
the UV) and reddening.  We therefore choose not to pursue the spectrum
of the UV excess further.


\section{The X-ray Spectra and Variability}

\subsection{X-ray Spectra Analysis}

\begin{table*}
\caption{Best-fitting parameters for Churazov fits to {\it Chandra} data.}
\label{XRayTable}
\begin{center}
\begin{tabular}{llllll}
\hline
\noalign{\smallskip}
Model   & \multicolumn{1}{c}{$N_H$}   &  \multicolumn{1}{c}{$\alpha$}  &  \multicolumn{1}{c}{$kT$}     & \multicolumn{1}{c}{$\chi^2/{\rm d.o.f.}$}   &  \multicolumn{1}{c}{0.3--7.0\,keV Flux} \\
        & \multicolumn{1}{c}{($10^{22}$\,cm$^{-2}$)} & & \multicolumn{1}{c}{(keV)}    &                         & \multicolumn{1}{c}{(erg\,cm$^{-2}$\,s$^{-1}$)} \\
\noalign{\smallskip}
\hline
\noalign{\smallskip}
\multicolumn{6}{l}{\em 2000 observation}\\
\noalign{\smallskip}
Power-Law      & $0.66 \pm 0.08$  & $1.83 \pm 0.15$   &  \ldots        & 1.00  & $1.37\times10^{-12}$   \\
Bremsstrahlung & $0.56^{+0.07}_{-0.06}$  & \ldots               &   $6.7^{+2.7}_{-1.6}$      & 1.01  & $1.37\times10^{-12}$   \\
Raymond-Smith  & $0.55 \pm 0.06$   & \ldots                &   $7.3^{+2.7}_{-1.6}$      & 1.05 & $1.53\times10^{-12}$   \\
\noalign{\smallskip}
\multicolumn{6}{l}{\em 2003 observation}\\
\noalign{\smallskip}
Power-Law      & $0.75^{+0.07}_{-0.08}$ & $2.17^{+0.12}_{-0.14}$ & \ldots & 0.93 & $2.91\times10^{-13}$ \\
Bremsstrahlung & $0.60\pm0.05$&  \ldots    & $3.97^{+0.7}_{-0.5}$ & 0.92 & $2.88\times10^{-13}$ \\
Raymond-Smith  & $0.65^{0.09}_{-0.07}$ & \ldots     & $3.29\pm0.4$ & 1.08 & $2.77\times10^{-13}$ \\
\noalign{\smallskip}
\hline
\end{tabular}
\end{center}
\end{table*}

We extracted spectra for both {\it Chandra} observations with CIAO
v3.2 and analysed them with XSPEC v12.2. We used standard CIAO
techniques to select response files according to the CCD
temperature. We considered many spectral models within XSPEC, all with
interstellar absorption, and report results from fitting with
power-law, bremsstrahlung, and Raymond-Smith models here. Blackbody
fits to the data were so poor that we neglect them for discussion. Due
to the lower count rate in the 2003 observation, we were prompted to
try using CASH statistics in tandem with the Churazov approximation
for chi-squared statistics (Cash, 1979; Churazov, 1996). For each set
of statistics to be used, we grouped the spectra into at least 15
counts per spectral bin; further binning results in a loss of spectral
information. Churazov weighting estimates the weight for a given
channel by averaging the counts in surrounding channels. CASH
statistics is a method designed to estimate the best-fit parameters
using unbinned or slightly binned data, which can be particularly
useful when the source yields few photons. We also checked the
consistency of CASH with unbinned data, and found the results to be
equivalent.

All models except the blackbody model give acceptable fits to both
data sets with the power-law providing the best fit. Both statistical
methods provide results that are equivalent to previously published
results by \citet{Kong:2002a} for the 2000 observation. The
best-fitting model is shown in Fig.~\ref{XSpecFig}. For the 2003
observation, we find similar results to \citet{Corbel:2008a}, and a
plot is also displayed in Fig.~\ref{XSpecFig}.  Table~\ref{XRayTable}
summarises the results of the spectral fitting.

\subsection{Color-Intensity Diagram}

\begin{figure}
 \begin{center}
  \includegraphics[angle=90,scale=0.35]{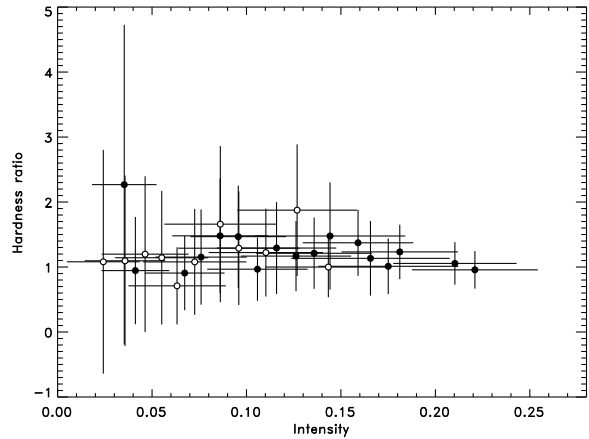}
  \caption{Hardness--intensity diagram from both {\it Chandra}
    observations. The two datasets were combined using the power-law
    model to estimate the change in sensitivity between the two
    epochs.  Solid circles indicate 2000 data and open circles that from 2003.}
\label{ColorIntensityFig}
 \end{center}
\end{figure}

To test for any variation in the spectrum with luminosity we construct
a hardness--intensity diagram.  We separated the data into two
bandpasses of approximately equal count rates, which were
0.3--1.75\,keV (soft) and 1.75--7.0\,keV (hard) for each set.  To
account for the reduction in sensitivity between the two epochs, we
used an assumed fixed model of a photo-absorbed power-law to calculate
the ratio of expected count rates in these bandpasses between the two
observations and hence rescale the 2000 observations to match those
from 2003. The resulting hardness--intensity diagram is shown in
Fig.~\ref{ColorIntensityFig}.  We see no variation in hardness either
within or between the two observations, the first of which had a
substantially higher count rate.  This indicates no detectable
spectral change over a factor of ten in source luminosity.

\section{Radio Spectra and Variability}
\label{RadioSection}

To constrain the radio spectral shape we first examine each dataset
individually.  We obtain $\alpha = -0.07\pm0.16$ (where $f_{\nu}
\propto \nu^{\alpha}$) from the VLA data and $\alpha = +0.04\pm0.38$
from the WSRT.  The difference between them is not statistically
significant so we also perform a joint power-law fit to both datasets,
allowing the relative normalisation of the two to vary as they sample
different portions of the lightcurve.  Primarily this is constrained
by the VLA data as the uncertainties in the WSRT fluxes are larger.
We find a spectral index of $\alpha=-0.05\pm0.15$.  All the fits are
consistent with the canonical flat spectrum as commonly seen in hard
state black hole binaries.

\begin{figure}
\includegraphics[angle=90,scale=0.35]{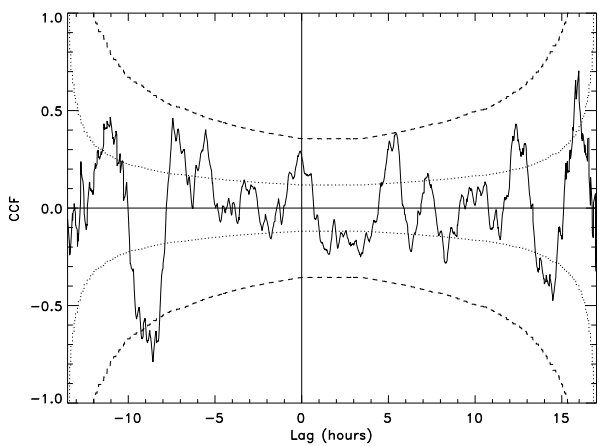}
\caption{Cross correlation function of radio lightcurve with respect
  to the X-rays.  Positive lags would indicate radio lagging behind
  the X-rays.  Dotted and dashed lines are $1-\sigma$ and $3-\sigma$
  expectations for uncorrelated variations.}
\label{CCFFig}
\end{figure}

The VLA data exhibit substantial variability.  We show the higher
quality 8.5\,GHz lightcurve together with that from {\it Chandra} in
Fig.~\ref{OptVarFig}.  The radio is clearly strongly variable and shows
both large flares and dips not dissimilar to those seen by {\it
  Chandra}.  The most dramatic feature rises from almost zero flux to
the highest peak in about 30\,min, as was also the case in the dataset
presented by \citet{MillerJones:2008a}.  No clear correlation between
radio and X-ray is apparent to the eye, however.  To further test
this, we calculate the cross-correlation function between the X-ray
and radio data and show this in Fig.~\ref{CCFFig}.  No compelling
correlation (or anti-correlation) is present although an unremarkable
peak is present at zero lag.

Since WSRT is a linear array, it covers a full synthesis in 12 hours
and hence is not capable of the high time-resolution achieved by the
VLA.  Nevertheless the source does show significant variability over
this period, even if a possible correlation with X-ray/optical is
difficult to infer. When the observation is split in 2 sub-intervals,
18:40--24:00 and 00:00--05:13, the resulting flux densities are: $S_1
= 0.25 \pm 0.04$\,mJy and $S_2 = 0.14 \pm 0.04$\,mJy respectively.
The second interval, covering D2--F2, overlaps with the first part of
the VLA observation, which shows a significant increase at 8.5\,GHz
beginning at 03:00 UT (with a peak flux density of 0.6\,mJy).  However, the
VLA flare only lasts for $\sim1$ hour, of which only about one-half
was covered by WSRT.  Finer sub-intervals proved impractical as there
was insufficient coverage of the UV plane to isolate flux from
\target.


\section{The Broad Band Spectral Energy Distribution}

\subsection{Comparison with existing data}

\begin{figure*}
\includegraphics[angle=90,scale=0.7]{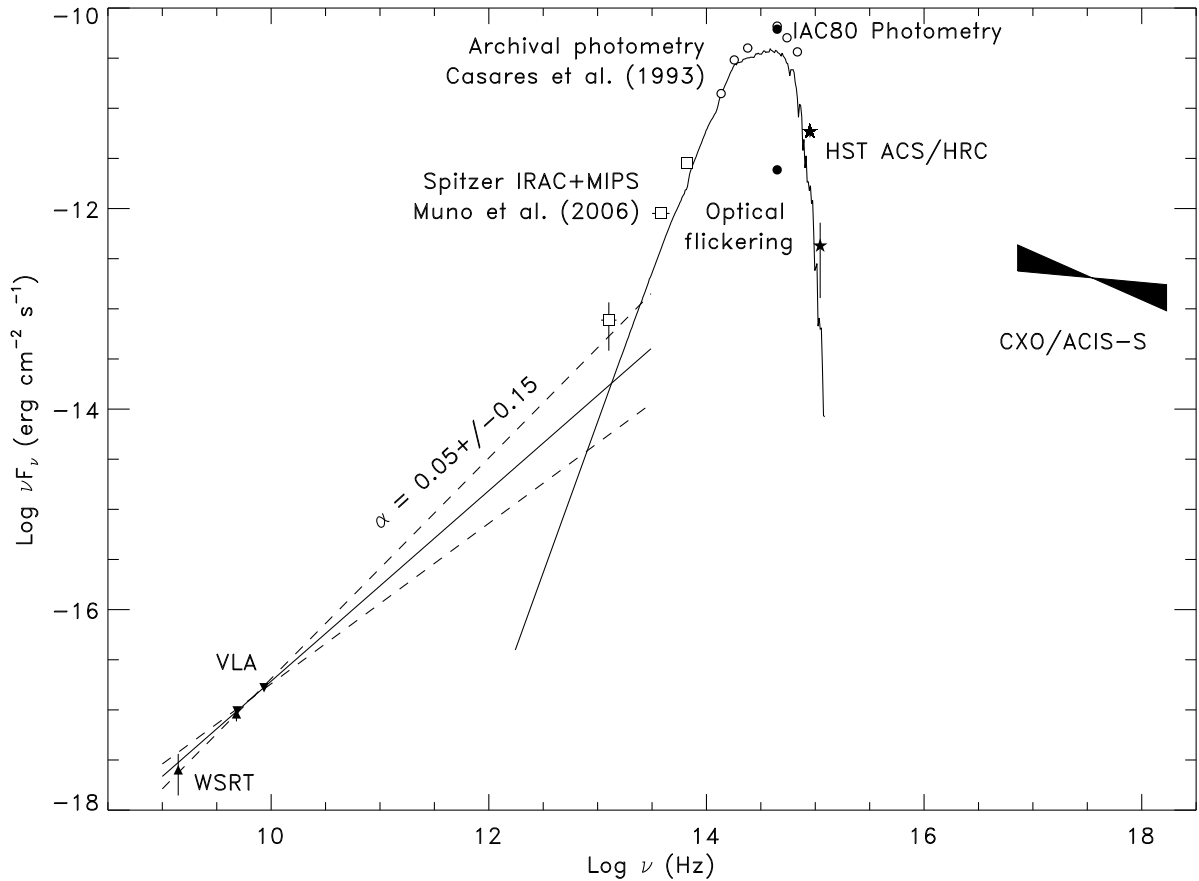}
\caption{Broad band SED.  Solid points are simultaneous data from our
  campaign, open points are non-simultaneous.  Triangles are used for
  radio data, squares for {\it Spitzer}, circles for optical
  photometry, and stars for {\it HST} UV photometry.  Solid lines
  indicate the stellar fit to the optical data and the power-law fit
  to the radio.}
\label{SEDFig}
\end{figure*}

Previous sections of this paper have focused primarily on one or two
bands; we will broaden our scope here to questions best addressed with
the whole SED.  We show the complete radio--X-ray SED in
Fig.~\ref{SEDFig}.  We have also compiled all the data points, as
plotted, in Table~\ref{DataTable} to allow easier comparison with
future models.  We include a point representing just the flickering
component in the optical to represent a lower limit on the accretion
light.  We adopt the extinction curve of \citet{Fitzpatrick:1999a} in
the optical and UV, \citet{Indebetouw:2005a} for the {\it
  Spitzer}/IRAC data and assume $A_{24\mu m} / A_{\rm K} \simeq 0.5$
following \citet{Chapman:2008a}.  Our dataset should supersede that
used by \citet{Narayan:1997a} for future modelling of \target.
Compared to earlier work we now have simultaneous X-ray and optical
data, shorter wavelength coverage extending into the near-UV, and
simultaneous radio observations.

\begin{figure}
\includegraphics[angle=90,scale=0.35]{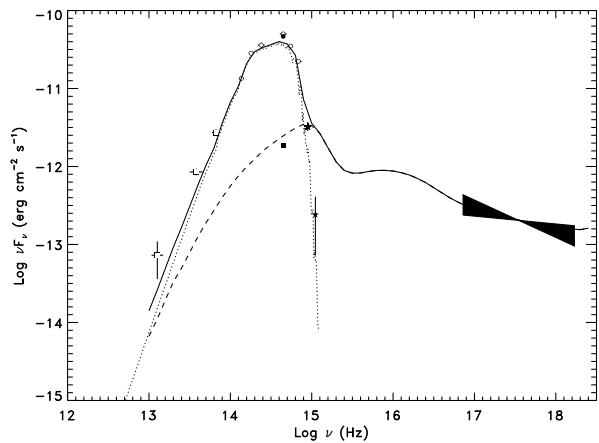}
\caption{Comparison of our SED with Model 1 of \citet{Narayan:1997a}.
  The dashed line is their ADAF model, the dotted line is now just the
  stellar emission.  The solid line is the composite of the two.}
\label{ADAFFig}
\end{figure}

In Fig.~\ref{ADAFFig} we illustrate the additional constraints now
possible, showing Model 1 of \citet{Narayan:1997a} overlaid on our SED
after scaling downwards to fit our X-ray spectrum.  This is of course
not intended to be a detailed model fit.  Not only should a new model
be computed to fit a lower X-ray luminosity, but there have been
refinements in modelling of advective accretion flows in the last
decade as well.  Our main point is to stress the strong constraint
provided by the low near-UV flux.  Model 1 was an acceptable fit to
the SED data available to \citet{Narayan:1997a} but clearly
over-predicts the near-UV substantially compared to our new data.
Alternative models with winds presented by \citet{Quataert:1999a}, on
the other hand, would be consistent with the UV data.  This constraint
on the UV to X-ray flux ratio can be used to constrain current and
future models.

\subsection{The origin of optical flickering}

The flux of the optical flickering component falls at an intriguing
location in the SED (Fig.~\ref{SEDFig}).  It is a little above a
straight extrapolation of either the X-ray or radio power-laws.  An
extrapolation of the radio is plausible, as a flat-spectrum is
expected to continue until a break to optically thin synchrotron at
higher frequencies.  We would not expect a straight extrapolation of
the X-ray spectrum at all, and all models for the broad-band SED
(e.g.\ \citealt{Narayan:1997a}) involve curvature between X-ray and
optical bands.)  Here correlations are more telling evidence than
fluxes.  We find an excellent correlation between X-ray and optical
continuum (Fig.~\ref{OptVarFig}), but no clear correlation between
X-ray and radio (and by extension optical).  This suggests that the
optical variability and the X-ray emission are closely connected,
either by direct emission from the same region, as expected from early
advection dominated accretion flow (ADAF) models, or indirectly
through X-ray irradiation of the disc, as inferred for H$\alpha$
emission by \citet{Hynes:2004a}.  The X-ray and radio emission, on the
other hand, do not obviously appear correlated on the timescales
observed (although a correlation on longer timescales is certainly
possible) suggesting that the mechanisms for variability in the two
are not connected.

\subsection{The extent of synchrotron emission}
\label{JetSection}

There is clearly a mid-IR excess prominent in the 24\,$\mu$m band.
While it is possible to partially explain this with emission from a
cool disc, synchrotron emission from the radio jet provides a better
description of the excess \citep{Gallo:2007a}.  An extrapolation of
the flat spectrum beyond the IR region is problematic for several
reasons.  We have argued above that the optical variability appears
directly or indirectly associated with the X-ray emitting component
rather than with the radio.  In addition, the low UV flux observed is
inconsistent with a straight extrapolation of the flat spectrum all
the way to the UV, and problematic even if the spectrum is assumed to
break in the optical.  This argues that a break in the jet spectrum at
longer wavelengths is needed so that the non-stellar optical/UV
emission originates from the disc rather than the jet.  The most
likely interpretation is then that the break from flat spectrum to
optically thin synchrotron occurs in the mid to far-IR, although given
the large amplitude of radio variability only simultaneous radio and
mid-IR observations could address this question conclusively.


\section{Conclusions}

We have compiled the comprehensive dataset describing the spectral
energy distribution of the black hole X-ray binary \target\ in
quiescence.  This includes new and simultaneous radio, optical, UV,
and X-ray data, supplemented with additional non-simultaneous data in
the optical through infrared.

Our specific findings are:

\begin{enumerate}

\item There is no need for a model to include a substantial
  contribution from accretion light over the near-UV to optical
  spectral range.  The SED over this range (including the vacuum UV)
  is well described by a K0\,{\sc IV} star reddened by $A_{\rm
    V}=4.0$.  The very low UV flux detected only marginally supports
  an accretion component, and rather tightly constrains some spectral
  models of quiescent accretion flows.

\item As suggested by \citet{Muno:2006a} no plausible companion
  spectrum and interstellar extinction can fully explain the mid-IR;
  an IR excess from a cool disc or possibly a jet
  \citep{Gallo:2007a} appears to be required.

\item The X-ray spectrum is consistent with a $\Gamma \sim 2$
  power-law as found by all other studies to date.  There is no
  evidence for any spectral variation over the range of a factor of 10
  in luminosity that has been observed over two epochs by Chandra.

\item The radio shows a flat spectrum in $f_{\nu}$ as is commonly seen
  in hard state X-ray binaries and associated with a compact jet.  The
  break frequency between a flat and optically thin spectrum most
  likely occurs in the sub-mm to far-IR regime, but is not
  conclusively constrained by existing data.  The radio is
  substantially variable but puzzlingly there is no clear correlation
  with X-ray variability.  If the radio variability reflects changes
  in the jet power, this suggests that the X-ray variations occur
  independently of changes in the jet, or with a very large lag (many
  hours) between the two.

\end{enumerate}

Overall, the likely interpretation of the SED is as follows:

\begin{enumerate}

\item The K0\,{\sc iii} companion star dominates the light from
  near-IR to near-UV wavelengths.

\item The cool outer regions of the accretion disc may produce a modest
  mid-IR excess seen by {\it Spitzer} \citep{Muno:2006a}.

\item The inner accretion flow produces highly variable X-ray emission
  but apparently only weak UV emission.  This irradiates the accretion
  disc and is responsible for emission line variability
  \citep{Hynes:2004a}.  The optical continuum variability may also
  arise from reprocessing but direct emission from the inner accretion
  flow remains a possibility.

\item A compact jet is driven from the inner accretion disc producing
  variable flat-spectrum radio emission likely extending to the far
  and possibly mid-IR.  The size of the jet is smaller than 4.5\,AU
  (about $30\times$ the binary separation) for a distance of 3.5\,kpc
  \citep{MillerJones:2008a}.

\end{enumerate}

\begin{table*}
\caption{Compilation of SED data used in Fig.~\ref{SEDFig}.}
\label{DataTable}
\begin{tabular}{lrrr}
 & \multicolumn{1}{c}{$\nu$} & \multicolumn{1}{c}{Observed $f_{\nu}$} & \multicolumn{1}{c}{Dereddened $f_{\nu}$} \\
 & \multicolumn{1}{c}{(Hz)}      & \multicolumn{1}{c}{(erg\,cm$^{-2}$\,s$^{-1}$\,Hz$^{-1}$)} & \multicolumn{1}{c}{(erg\,cm$^{-2}$\,s$^{-1}$\,Hz$^{-1}$)}  \\
\noalign{\smallskip}
\hline
\noalign{\smallskip}
\multicolumn{3}{l}{\em Simultaneous Observations} \\
WSRT (1.4\,GHz) & $1.40\times10^{9}$ & $(1.8\pm0.8)\times10^{-27}$ \\
WSRT (4.8\,GHz) & $4.80\times10^{9}$ & $(1.9\pm0.3)\times10^{-27}$ \\
VLA (4.9\,GHz) & $4.86\times10^{9}$ & $(2.01\pm0.10)\times10^{-27}$ \\
VLA (8.6\,GHz) & $8.60\times10^{9}$ & $(1.93\pm0.10)\times10^{-27}$ \\
IAC80 (R) & $4.55\times10^{14}$ & $(5.62\pm0.05)\times10^{-27}$ & $(9.37\pm0.09)\times10^{-26}$ \\
HST (F330W) &  $8.93\times10^{14}$ & $(8.9\pm1.1)\times10^{-30}$ & $(3.6\pm0.5)\times10^{-27}$ \\
HST (F250W) & $1.11\times10^{15}$ & $(7.0\pm4.9)\times10^{-31}$ & $(2.2\pm1.5)\times10^{-28}$ \\
Chandra & $8.81\times10^{17}$ & $2.12\times10^{-31}$ \\
\noalign{\smallskip}
\multicolumn{3}{l}{\em Non-simultaneous Observations} \\
Spitzer (24\,$\mu$m)  & $1.27\times10^{13}$ & $(0.44\pm0.22)\times10^{-26}$ & $(0.57\pm0.27)\times10^{-26}$ \\
Spitzer (8\,$\mu$m)   & $3.81\times10^{13}$ & $(1.76\pm0.18)\times10^{-26}$ & $(2.23\pm0.22)\times10^{-26}$ \\
Spitzer (4.5\,$\mu$m) & $6.68\times10^{13}$ & $(3.22\pm0.32)\times10^{-26}$ & $(4.08\pm0.38)\times10^{-26}$ \\
Archival $K$ & $1.36\times10^{14}$ & $(6.4\pm0.3)\times10^{-26}$ & $(9.9\pm0.5)\times10^{-26}$ \\
Archival $H$ & $1.80\times10^{14}$ & $(7.7\pm0.4)\times10^{-26}$ & $(15.7\pm0.9)\times10^{-26}$ \\
Archival $J$ & $2.39\times10^{14}$ & $(5.3\pm0.4)\times10^{-26}$ & $(15.0\pm1.1)\times10^{-26}$ \\
Archival $R$ & $4.47\times10^{14}$ & $(6.86\pm0.06)\times10^{-27}$ & $(11.21\pm0.10)\times10^{-26}$ \\
Archival $V$ & $5.50\times10^{14}$ & $(1.54\pm0.03)\times10^{-27}$ & $(6.35\pm0.12)\times10^{-26}$ \\
Archival $B$ & $6.85\times10^{14}$ & $(2.25\pm0.10)\times10^{-28}$ & $(3.26\pm0.15)\times10^{-26}$ \\
\noalign{\smallskip}
\hline
\end{tabular}
\end{table*}


\section*{Acknowledgements}

We are grateful to Valerie Mikles for many helpful comments on this
manuscript.  Chandra observations were supported by NASA grant
GO3-4044X.  C.~B. acknowledges support from a LaSpace GSRA award.
E. G. acknowledges support from a Hubble Fellowship.  The WHT is
operated on La Palma by the ING in the Spanish Observatorio del Roque
de los Muchachos of the Instituto de Astrof\'\i{}sica de Canarias.
Gemini observations were obtained under program GN-2003A-Q-12 by the
Gemini Observatory which is operated by the Association of
Universities for Research in Astronomy, Inc., under a cooperative
agreement with the NSF on behalf of the Gemini partnership: the
National Science Foundation (United States), the Science and
Technology Facilities Council (United Kingdom), the National Research
Council (Canada), CONICYT (Chile), the Australian Research Council
(Australia), Minist\'{e}rio da Ci\^{e}ncia e Tecnologia (Brazil) and
SECYT (Argentina)

\label{lastpage}

\end{document}